%% file: paper.tex
\documentclass[sigconf, nonacm]{acmart}

\usepackage[a-2b]{pdfx}
\usepackage[T1]{fontenc}
\usepackage{etoolbox,graphicx,setspace,listings,multicol,xspace,enumitem,booktabs,xcolor}
\usepackage{balance}
\usepackage{subcaption}
\usepackage{algorithm,mathtools}
\usepackage[noend]{algpseudocode}
\usepackage[skip=10pt,labelfont=bf]{caption}
\usepackage{tabularx}
\usepackage{amsmath}
\usepackage{dsfont}
\usepackage{accents}

\newcommand\vldbdoi{10.14778/3476249.3476285}
\newcommand\vldbpages{2341 - 2354}
\newcommand\vldbvolume{14}
\newcommand\vldbissue{11}
\newcommand\vldbyear{2021}
\newcommand\vldbauthors{\authors}
\newcommand\vldbtitle{\shorttitle} 
\newcommand\vldbpagestyle{empty}

\newcommand*\LSTfont{\Small\ttfamily\SetTracking{encoding=*}{-60}\lsstyle}
\input{macros}

\begin{document}

\title{Accelerating Approximate Aggregation Queries with Expensive Predicates}
\author{Daniel Kang*}
\affiliation{
  \institution{Stanford University}
}
\email{ddkang@stanford.edu}
\author{John Guibas*}
\affiliation{
  \institution{Stanford University}
}
\email{jtguibas@stanford.edu}
\author{Peter Bailis}
\affiliation{
  \institution{Stanford University}
}
\email{pbailis@stanford.edu}
\author{Tatsunori Hashimoto}
\affiliation{
  \institution{Stanford University}
}
\email{thashim@stanford.edu}
\author{Yi Sun}
\affiliation{
  \institution{University of Chicago}
}
\email{yi.sun@uchicago.edu}
\author{Matei Zaharia}
\affiliation{
  \institution{Stanford University}
}
\email{matei@cs.stanford.edu}

\input{tex/abstract}

\maketitle

\pagestyle{\vldbpagestyle}
\begingroup\small\noindent\raggedright\textbf{PVLDB Reference Format:}\\
\vldbauthors. \vldbtitle. PVLDB, \vldbvolume(\vldbissue): \vldbpages, \vldbyear.\\
\href{https://doi.org/\vldbdoi}{doi:\vldbdoi}
\endgroup
\begingroup
\renewcommand\thefootnote{}\footnote{\noindent
* Marked authors contributed equally.

This work is licensed under the Creative Commons BY-NC-ND 4.0 International License. Visit \url{https://creativecommons.org/licenses/by-nc-nd/4.0/} to view a copy of this license. For any use beyond those covered by this license, obtain permission by emailing \href{mailto:info@vldb.org}{info@vldb.org}. Copyright is held by the owner/author(s). Publication rights licensed to the VLDB Endowment. \\
\raggedright Proceedings of the VLDB Endowment, Vol. \vldbvolume, No. \vldbissue\ %
ISSN 2150-8097. \\
\href{https://doi.org/\vldbdoi}{doi:\vldbdoi} \\
}\addtocounter{footnote}{-1}\endgroup

\input{tex/intro}
\input{tex/overview}
\input{tex/query_proc}
\input{tex/analysis}

\input{tex/eval}

\input{tex/rel_work}
\input{tex/conclusion}

\begin{acks}
{
This research was supported in part by affiliate members and other supporters of
the Stanford DAWN project---Ant Financial, Facebook, Google, Infosys, NEC, and
VMware---as well as Toyota Research Institute (``TRI''), Northrop Grumman, Amazon Web
Services, Cisco, and the NSF under CAREER grant CNS-1651570. Any opinions,
findings, and conclusions or recommendations expressed in this material are
those of the authors and do not necessarily reflect the views of the NSF. TRI provided funds to assist the authors with their
research but this article solely reflects the opinions and conclusions of its
authors and not TRI or any other Toyota entity.
}
\end{acks}

\bibliographystyle{ACM-Reference-Format}
\bibliography{paper}

\end{document}

%% file: macros.tex
\newcommand*{\algname}{\textsc{ABae}\xspace}
\newcommand*{\algmulti}{\textsc{ABae-MultiPred}\xspace}
\newcommand*{\alggroup}{\textsc{ABae-GroupBy}\xspace}

\newcommand*{\noscope}{\textsc{NoScope}\xspace}
\newcommand*{\blazeit}{\textsc{BlazeIt}\xspace}

\newcommand{\minihead}[1]{{\vspace{.45em}\noindent\textbf{#1.} }}

\newcommand{\ubar}[1]{\underaccent{\bar}{#1}}

\newcommand*{\universe}{\mathcal{D}}
\newcommand*{\pall}{ {p_{\textrm{all}}} }

\newcommand*{\Exp}{\mathbb{E}}
\newcommand*{\Var}{\text{Var}}
\newcommand{\floor}[1]{\lfloor #1 \rfloor}

\newcommand*{\hmuall}{\hat{\mu}_{\text{all}}}
\newcommand*{\muall}{\mu_{\text{all}}}

\newtheorem{prop}{Proposition}
\newenvironment{proofsketch}{%
  \renewcommand{\proofname}{Proof sketch}\proof}{\endproof}

\newtoggle{rcolors}
\toggletrue{rcolors}

%% file: tex/abstract.tex
\begin{abstract}

Researchers and industry analysts are increasingly interested in computing
aggregation queries over large, unstructured datasets with selective predicates
that are computed using expensive deep neural networks (DNNs). As these DNNs are
expensive and because many applications can tolerate approximate answers,
analysts are interested in accelerating these queries via approximations.
Unfortunately, standard approximate query processing techniques to accelerate
such queries are not applicable because they assume the result of the
predicates are available ahead of time. Furthermore, recent work using cheap approximations (i.e., proxies) do
not support aggregation queries with predicates.

To accelerate aggregation queries with expensive predicates, we develop and
analyze a
query processing algorithm that leverages proxies (\algname). \algname must
account for the key challenge that it may sample records that do not satisfy the
predicate. To address this challenge, we first use the proxy to group records
into strata so that records satisfying the predicate are ideally grouped into
few strata.
Given these strata, \algname uses pilot sampling and plugin estimates to
sample according to the optimal allocation.
We show that \algname converges at an optimal rate
in a novel analysis of stratified sampling with draws that may not satisfy the
predicate. We further show that \algname outperforms on baselines on six
real-world datasets, reducing labeling costs by up to 2.3$\times$.


\end{abstract}

%% file: tex/intro.tex
\section{Introduction}
\label{sec:intro}

Analysts are interested in computing statistics over large, unstructured
datasets where only a fraction of the data is of interest (i.e., with a
selective predicate) with low computational cost. Machine learning (ML) methods
are increasingly used to automatically answer such queries. For example, a media
studies researcher may be interested in computing the average viewership (the
statistic) of presidential candidates on TV news (the predicate)
\cite{hong2020analyzing}. To answer such queries, the researcher may deploy an
expensive face detection deep neural network (DNN) to find all faces in the
dataset and filter by presidential candidates, e.g.,
\begin{lstlisting}
SELECT AVG(views) FROM video
WHERE contains_candidate(frame, 'Biden')
\end{lstlisting}
Critically, these DNNs can be expensive to execute. For example,
executing a state-of-the-art face detection DNN on the past year of MSNBC News
would cost \$262,000 on cloud compute infrastructure (NVIDIA V100, Amazon Web
Services)~\cite{zhang2018face}. Due to limited computational budgets, many
organizations cannot exhaustively execute these expensive ML methods over the
entirety of the dataset.

Fortunately, many applications can tolerate approximations (as is standard in
the approximate query processing (AQP) literature~\cite{li2018approximate}) so
answering queries does not require exhaustively executing the expensive DNN. As
is standard in AQP, a key requirement with approximate answers are statistical
guarantees on query results. For example, the media studies researcher may
require such guarantees to make precise claims about bias in TV news.
Furthermore, these requirements are standard in scientific analyses. As such, we
focus on queries with statistical guarantees in this work.

Unfortunately, standard techniques in AQP, ranging from histograms
\cite{piatetsky1984accurate}, sketches~\cite{braverman2013generalizing}, and
others~\cite{agarwal2013blinkdb}, assume that the fields used in the predicates
are already available, i.e., as structured records in a database. In contrast,
we cannot precompute results as an expensive ML method is required to compute
the predicate in our setting, e.g., we would have to execute an expensive face
detector on every video frame to answer the query above. Recent work has focused
on using cheap approximations (i.e., proxy models) to accelerate queries without
having to pre-compute expensive DNNs~\cite{kang2017noscope, kang2019blazeit,
kang2020approximate, hsieh2018focus, anderson2018predicate}.  For example, a
proxy for presidential candidates might be a cheap classifier in contrast to a
full object detection DNN.  Unfortunately, existing work either does not provide
statistical guarantees on query accuracy (e.g., \noscope~\cite{kang2017noscope},
Focus~\cite{hsieh2018focus}, Tahoma~\cite{anderson2018predicate}) or accelerates
other query types (e.g., selection queries~\cite{kang2020approximate}, aggregation
queries without predicates~\cite{kang2019blazeit}, and limit
queries~\cite{kang2019blazeit}).

We propose and analyze \algname (\textbf{A}ggregation with Expensive
\textbf{B}in\textbf{A}ry Pr\textbf{E}dicates), a query processing algorithm leveraging
stratified and pilot sampling \cite{kish1965survey} to accelerate
linear aggregation queries (\texttt{SUM}, \texttt{COUNT}, and
\texttt{AVG}) with expensive predicates and statistical guarantees
on query accuracy. We further extend \algname to support common aggregation
patterns, including queries with multiple predicates and with group by keys.

\algname leverages two key opportunities to accelerate such queries: proxy
models and stratified sampling. That is, \algname splits the dataset into
disjoint groups (strata), samples within strata, and computes a weighted
average to obtain the final answer. \algname must account for three
key challenges, as the predicate results are not available ahead
of time: 1) strata selection, 2) budget allocation between strata, and 3)
stochastic draws (i.e., sampling a record that may not match the predicate). We
provide a principled stratification approach,
leverage pilot sampling for budget allocation \cite{kish1965survey},
and provide a novel analysis of stratified sampling with stochastic draws
that shows that \algname converges at an optimal rate.

To address strata selection, \algname uses the proxy model. We
assume the proxy provides information about the likelihood of a record
satisfying the predicate \cite{kang2017noscope, kang2020approximate,
anderson2018predicate, hsieh2018focus}. Since the proxy does not give
information about the statistic, we stratify records by proxy score quantile.
Under a mild monotonicity assumption on the
proxy~\cite{guo2017calibration}, this stratification will group records that are
approximately equally likely to match the predicate in the same stratum.
Intuitively, if the proxy is perfect (i.e., matches the predicate) and is
independent of the statistic, this stratification will minimize the sampling
variance. While \algname performs best when given proxy
models which approximate the expensive predicate well, \algname still returns
correct answers regardless of proxy model quality.

Given a stratification, our analysis shows that the optimal allocation depends
on two key, per-strata quantities: the fraction of records that match the
predicate ($p_k$) and the standard deviation of the statistic within a stratum
($\sigma_k$). Concretely, the optimal allocation is proportional to $\sqrt{p_k}
\sigma_k$. However, we do not know these quantities ahead of time.

\algname proceeds in two stages to address this challenge. First, \algname will
estimate $p_k$ and $\sigma_k$ using a fraction of the total sampling budget.
Then, \algname will allocate the sampling budget using our plug-in estimates of
$p_k$ and $\sigma_k$. We prove that \algname's algorithm matches the expected
error rates of the optimal stratified sampling allocation given the key
quantities. Finally, to provide confidence intervals, \algname uses a
bootstrapping procedure which only adds minimal computational overhead.

We also extend \algname to support group bys (\alggroup) and complex expressions
involving multiple Boolean predicates (\algmulti). To support group by
statements, we adapt our sample allocation strategy to minimize the maximum of
the expected mean squared error of the groups (minimax error). We show a
numerical optimization procedure can recover the optimal allocation for the
minimax error.  We also support combining multiple expensive predicates and
their respective proxy models through negations, conjunctions, and disjunctions.

Finally, a key challenge in leveraging proxy models is to ensure efficient query
answers despite potentially poor proxy model quality. Because \algname
always produces valid results, we only need to address efficiency. To address
this challenge, we derive a formula which computes the relative gain of using a
given proxy. Then, by using a cheap procedure which can
estimate the quantities in the formula, \algname can calculate expected
performance gains of proxy models and select the best proxy model at query time.

We evaluate \algname and its extensions on six real-world datasets spanning
text, images, and video. We show that \algname outperforms uniform sampling by
up to 2.3$\times$. We also
provide experiments to show that our methods for creating confidence intervals,
executing group by
aggregation queries, and forming complex predicates from multiple proxy models
outperforms baselines.

In summary, our contributions are:
\begin{enumerate}
  \item We develop \algname and its extensions, \algmulti and \alggroup, to
  accelerate aggregation queries with expensive predicates via proxy models.

  \item We provide a theoretical analysis of \algname and show that it matches the
  expected error of the optimal stratified sampling algorithm asymptotically.

  \item We evaluate these techniques on text, image, and video datasets, showing
  that \algname significantly outperforms uniform sampling.

\end{enumerate}

%% file: tex/overview.tex
\section{Overview and Query Semantics}
\label{sec:query}

\subsection{Overview}
\minihead{Target setting}
\algname targets aggregation queries that contain one or more predicates that
are expensive to evaluate. These predicates typically require executing
expensive DNNs or querying human labelers. We assume the statistic can be
computed in conjunction with the predicates or is cheap to compute. We
support aggregation queries targeting \texttt{AVG}, \texttt{SUM}, and
\texttt{COUNT} statistics. We do not support other aggregation types, such as
\texttt{COUNT DISTINCT} or \texttt{MAX}.


\minihead{Proxies}
We further assume access to a proxy model per predicate, which returns a
continuous value between $0$ and $1$. While not necessary for correctness, high
quality proxies will return scores that are correlated with the predicate.
These proxies can be orders of magnitude cheaper than oracles (e.g., over 4,000
images/second for the proxy vs 3 fps for the oracle~\cite{kang2021jointly}).
Thus, as is standard in the literature, we assume these proxies are
substantially cheaper than the oracle methods so the proxies can be exhaustively
executed over the entire dataset~\cite{kang2017noscope, canel2019scaling,
kang2019blazeit, xu2019vstore}.


\subsection{Examples}
\label{sec:examples}
\minihead{TV news}
Consider a media studies researcher studying how the presence of presidential candidate
affects viewership. The researcher is willing to query the expensive DNN at most
10,000 times and computes the average viewership with
the following:
\begin{lstlisting}
SELECT AVG(views) FROM news
WHERE contains_candidate(frame, 'Biden')
ORACLE LIMIT 10,000 USING proxy(frame)
WITH PROBABILITY 0.95
\end{lstlisting}
where \texttt{contains\_candidate} is computed via a face detection DNN and the
proxy may be trained via specialization~\cite{kang2017noscope}.


\minihead{Traffic analysis}
Consider an urban planner studying traffic patterns. The planner is interested
in understanding waiting times at traffic lights and executes the following
query
\begin{lstlisting}
SELECT AVG(count_cars(frame)) FROM video
WHERE count_cars(frame) > 0
  AND red_light(frame)
ORACLE LIMIT 1,000 USING proxy(frame)
WITH PROBABILITY 0.95
\end{lstlisting}
where \texttt{count\_cars} is computed via an object detection DNN and
\texttt{red\_light} is computed by a human labeler. The proxy could be computed
via an embedding index for unstructured data~\cite{kang2020task}.

\minihead{Analyzing historical newspaper scans}
Consider political scientists that are interested in computing statistics (e.g.,
fraction of articles with positive sentiment) over editorials (i.e., the
predicate) in historical newspaper scans. Computing these statistics requires
executing expensive OCR and text processing DNNs.

\begin{figure}[t!]
\begin{lstlisting}[frame=single]
SELECT {AVG | SUM | COUNT} ({field | EXPR(field)})
  FROM table_name WHERE filter_predicate
  [GROUP BY key]
  ORACLE LIMIT o USING proxy
  WITH PROBABILITY p
\end{lstlisting}
\caption{
Syntax for \algname.  Users provide a statistic to compute, an expensive
predicate, an oracle limit, proxy scores, and a success probability. As is
standard for aggregation queries, users may specify a group by key.
}
\label{fig:syntax}
\end{figure}

\subsection{Query Syntax and Semantics}
We show the query syntax for \algname in Figure~\ref{fig:syntax}. As with
standard AQP systems, \algname accepts a sampling budget and a probability of
error and will return an approximate answer to the query and a confidence
interval (CI). Our CI semantics are the standard frequentist CI semantics
provided by other AQP systems~\cite{agarwal2013blinkdb}. In particular,
our CI semantics are valid regardless of proxy quality.

In contrast to standard AQP systems, \algname assumes that the predicate is
expensive to evaluate. We refer to the methods to execute the predicates as
``oracles'' \cite{kang2020approximate, kang2019blazeit}.
These oracles typically involve executing an expensive DNN and post-processing
the result, e.g., executing Mask R-CNN to extract object types and positions
from a frames of video and filtering by frames that contain at least two cars.
Other use cases may require a human labeler. We further assume that the
statistic is either cheap to compute or can be extracted by post-processing the
oracle results.

To accelerate these queries, the user also provides a proxy function that
computes per-record proxy scores for each predicate. These proxy scores are
ideally correlated with the result of the predicate and substantially cheaper
than the oracle predicates. Nonetheless, our algorithms will provide valid
results even if the proxy scores are of poor quality: proxy correlation
will only affect performance, not correctness.

\algname aims to return query results that minimize the mean squared error (MSE)
between the approximate result and the result when exhaustively executing the
query. \algname further aims to return CIs that are as tight as
possible while maintaining the probability of success.

\subsection{Query Formalism}
\label{sec:query-formalism}

Formally, let $\universe = \{ x_i \}$ be the set of data records, $O(x) \in \{0,
1\}$ be the oracle predicate, and $X_i = f(x_i) \in \mathbb{R}$ be the
expression the query aggregates over. Let $\universe^+ = \{x \in \universe :
O(x) = 1 \}$.  Finally, let $N$ be the sample budget.

\algname computes $\mu = \sum_{x \in \universe^+} f(x) / |\universe^+|$ via an
approximation, $\hat{\mu}$, with a fixed sampling budget $N$. We measure query
result quality by the MSE, i.e., $|\mu - \hat{\mu}|^2$. \algname returns a
CI $[\ubar{\mu}, \bar{\mu}]$. \algname further aims
to minimize the length of the CI $\bar{\mu} - \ubar{\mu}$
subject to $\mu \in [\ubar{\mu}, \bar{\mu}]$ with the specified probability and
sample budget, over randomizations of the query procedure.

%% file: tex/query_proc.tex
\section{Algorithm Description and Query Processing}
\label{sec:query-proc}

We describe \algname for accelerating aggregation queries with expensive
predicates. We first describe accelerating queries with a single predicate. We
then describe three natural extensions: queries with a group by key, queries
with multiple predicates, and estimating proxy quality.

\begin{table}[t!]
\centering
\caption{Summary of notation.}
\label{table:notation}
\small
\begin{tabular}{ll}
  Symbol & Description \\ \hline
  $\universe$   & Universe of data records \\
  $\mathcal{S}$           & Stratification, i.e., $k$ strata \\
  $\mathcal{P}(x)$ & Proxy model \\
  $N$           & User-specified sampling budget \\
  $K$           & Number of strata \\
  $\mathcal{O}(x)$        & Oracle predicate \\
  $X_{k, i}$    & $i$th sample from stratum $k$ \\
  $p_k$         & Predicate positive rate \\
  $\pall$       & $\sum_k p_k$ \\
  $w_k$         & Normalized $p_k$, i.e., $p_k / \pall$ \\
  $\mu_k$       & $\Exp[X_{i, k}]$ \\
  $\muall$      & $\sum p_k \mu_k / \pall$ \\
  $\sigma_k^2$  & $Var[X_{i, k}]$ \\
  $N_1$         & Number of samples in Stage 1 \\
  $N_2$         & Number of samples in Stage 2
\end{tabular}
\end{table}

\subsection{\algname with a Single Predicate}

\minihead{Overview}
\algname leverages stratified sampling and pilot sampling \cite{kish1965survey} to accelerate aggregation queries with
expensive predicates. Namely, \algname splits the dataset into disjoint subsets
called strata. Then, \algname allocates sampling budget to the strata and
combines the per-strata estimates to give the final estimate.

Our setting involves three distinct challenges. First, since not all records
satisfy the predicate, we may not sample a valid record. This change, while
seemingly simple, changes the optimal allocation and requires new theoretical
analysis to prove convergence rates. Second, we must construct the strata
without knowing which records satisfy the predicate. Third, we do not know $p_k$
(the predicate positive rate) and $\sigma_k$ (the standard deviation), which are
necessary for computing the optimal allocation.

To address these issues, we leverage a two-stage sampling algorithm.
\algname first estimates the key quantities necessary
for optimal allocation: $p_k$ and $\sigma_k$ (also known as pilot
sampling). \algname then uses these estimates
to allocate sampling budget in the Stage 2. We show in
Section~\ref{sec:analysis} that \algname achieves an optimal rate.

\input{tex/main_alg.tex}

\minihead{Formal description}
Recall that $p_k$ is the predicate positive rate and that $\sigma_k^2$ is the
variance of the statistic. Furthermore, recall that $\universe$ is the full
dataset, $O(x)$ is the oracle predicate, and $X_i$ are the samples. Denote
$X_{k, i}$ to be the $i$th \emph{positive} sample from stratum $k$.

Additionally, denote $K$ to be the number of strata, $N_1$ to be the number of
samples in Stage 1, and $N_2$ to be the number of samples in Stage 2, which are
parameters to \algname. \algname will compute several other quantities,
including $\pall = \sum_k p_k$, $w_k = p_k / \pall$ the normalized $p_k$, and
$\mu_k = \Exp[X_{k, i}]$ the per stratum mean. We summarize the notation in
Table~\ref{table:notation}.

We present the pseudocode for the sampling algorithm in
Algorithm~\ref{alg:full-alg}. \algname creates the strata by ordering the
records by proxy score and splitting into $K$ strata by quantile.

\algname will then perform a two-stage sampling procedure. In Stage 1, \algname
samples $N_1$ samples from each of the $K$ strata to estimate $p_k$ and
$\sigma_k$, which are the key quantities for determining optimal allocation. In
Stage 2, \algname will allocate the remaining samples proportional to our
estimates of the optimal allocation.

\algname construct plugin estimates for $p_k$ and $\mu_k$, denoted $\hat{p}_k$
and $\hat{\mu}_k$ respectively. To compute its final
estimates, \algname will use all the samples from Stage 1 and Stage 2 to
compute $\hat{p}_k$ and $\hat{\mu}_k$. \algname will return the estimate
$\sum_k \hat{p}_k \hat{\mu}_k / \sum_k \hat{p}_k $ as the approximate answer.

As the final estimates are sensitive to the estimate of $p_k$, i.e.,
$\hat{p}_k$, we find that reusing samples between stages dramatically improves
performance (Section~\ref{sec:eval-sensitivity}).

We defer the proofs of convergence and rates to Section~\ref{sec:analysis}.

\minihead{Confidence intervals}
We use the non-parametric bootstrap~\cite{efron1994introduction} to compute
confidence intervals, which resamples existing samples. Since the per-stratum
samples from both stages of \algname are independent and identically
distributed (i.i.d.), we resample from samples across both stages.

\input{tex/bootstrap_alg.tex}

We present the pseudocode for the bootstrap procedure in
Algorithm~\ref{alg:bootstrap-alg}. \algname bootstraps across both stages of the
sampling algorithm to form CIs. We formally show the
validity of the bootstrap in an extended technical report~\cite{kang2021proof}.
We further show that our procedure produces CIs that are
nominally correct in Section~\ref{sec:eval}.

In standard AQP, the bootstrap is considered an expensive procedure as it
requires resampling and recomputing the statistic. However, in our setting, we
assume that the oracle predicate is expensive to execute. As a result, the
bootstrap is computationally cheap compared to the cost of obtaining the
samples. Concretely, in several of our experiments, executing 1,000
bootstrap trials using unoptimized Python code on a single CPU core is as
expensive as executing 2,500 oracle calls on an NVIDIA T4 accelerator, which
corresponds to under 0.3\% of a medium-sized dataset.

\minihead{Setting parameters}
\algname requires setting two parameters: the fraction of samples between Stage
1 and 2 and the number of strata. We recommend using 30-50\% of samples in Stage 1
and $K$ to be maximal such that every strata receives at least 100 samples in
Stage 1. Our experiments show that \algname is not sensitive to these
parameters, but that these settings tend to do best
(Section~\ref{sec:eval-sensitivity}).

\subsection{Group Bys}
\label{sec:groupby-desc}
We extend \algname to support queries with group by keys (\alggroup). We focus
on minimizing the maximum
error (minimax error) over groups. Other objectives can be supported (e.g.,
the sum of errors), but we defer optimizing other objectives to future work.
%
As an example of a query with a group by key, consider:
\begin{lstlisting}
SELECT COUNT(frame), person FROM VIDEO
WHERE person IN ('Biden', 'Trump')
GROUP BY person
\end{lstlisting}
where, for simplicity, we assume that person is a virtual field extracted by an
expensive face detection DNN.

We assume that the group by key is expensive to compute and consider two
different scenarios. In the first scenario, a single oracle determines the group
key directly. For example, in the query above, a single oracle would directly
classify a person as Biden or Trump. In the second scenario, there is an oracle
\emph{per group}, where each oracle can classify whether a record belongs to a
specific group key or not. For example, in the query above, we must execute two
oracles: one to classify whether or not a person is Biden and another for Trump.
We have found this scenario to be common when practitioners do not build their
own models. We separate these two cases as they have different optimization
objectives.

To optimize the minimax error, we formulate the allocation of the samples to
minimize the minimax error as a non-linear optimization problem. Concretely,
consider the multiple oracle setting. Given a stratification of group $g$ (with
a total of $G$ groups), denote the error of this stratification as
$\textrm{Err}(g)/N$, for $N$ samples. Namely, $\textrm{Err}(g)/N$ is equivalent
to the error when using a single predicate. Then, we aim to optimize the
following objective
\begin{align}
  \mathcal{L} = \min_{\Lambda \in [0, 1]^G, \sum_{l=1}^G \Lambda_l = 1}
  \left(
    \max_{g} \frac{\textrm{Err}(g)}{\Lambda_l N}
  \right)
  \label{eq:groupby-minimax}
\end{align}
where $\Lambda$ is a weight vector corresponding to the sample allocation
between groups. We show that this optimization problem can be numerically
optimized using the Nelder-Mead simplex algorithm~\cite{sorensen1982newton}.

Concretely, \alggroup proceeds as follows:
\begin{enumerate}
  \item Sample uniformly at random to estimate the quantities needed to compute
  $\textrm{Err}(g)$.
  \item Solve Eq.~\ref{eq:groupby-minimax} via the Nelder-Mead simplex
  algorithm~\cite{sorensen1982newton}.
  \item Sample according to the allocation in the previous step.
  \item Return the combined estimates.
\end{enumerate}
We defer the full formulation to Section~\ref{sec:group-by-analysis}.


\subsection{Complex Predicates}
In addition to a queries with a single predicate, we extend \algname to support
queries that contain any number of conjunctions, disjunctions, and negations
(\algmulti). Presently, we assume that \algmulti receives as input a set of
per-record proxy scores per predicate. We show an example of such a query in
Section~\ref{sec:examples}.

To answer such queries, \algmulti will combine the proxy scores from the predicates to
obtain a single, per-record set of scores that ideally indicates the likelihood
of a record matching the whole expression. \algmulti combines the proxy scores by
transforming the expression into an arithmetic expression with the following
substitutions:
\begin{enumerate}
  \item Negations are replaced by subtraction from one.
  \item Conjunctions are replaced by products.
  \item Disjunctions are replaced by max.
\end{enumerate}

\algmulti's approach will return exact results if the proxies are perfectly
calibrated and perfectly sharp. While this assumption does not hold in practice,
we show that our approach works well in practice in Section~\ref{sec:eval}.

\subsection{Selecting Proxies}
In some applications, users may have to select between several viable proxies
for an expensive predicate. For example, suppose the user wishes to filter
emails by spam. The user can provide several rule-based proxies in the form of
detecting keywords, such as ``money'' or ``\$''.

A key question is which proxy will provide the lowest MSE for a given budget.
To estimate performance improvements, \algname will use the samples from Stage
1. For each proxy, \algname will construct the strata and estimate the
corresponding $p_k$ and $\sigma_k$. \algname will use the MSE formula for the
perfect information, deterministic draws setting to estimate the optimal
achievable MSE (Proposition~\ref{prop:opt-mse-deterministic}). Given these
estimates, \algname will take the top proxy as the proxy to use in the query. We
note that this procedure can reuse samples and only adds negligible
computational overhead.

Although the formula for the perfect information, deterministic draws setting
does not directly apply, we find it is a good predictor of relative performance.

Finally, \algname can combine proxies by sampling randomly in Stage 1 and using
these samples to train a logistic regression model using the proxies as features
and the predicate as the target.

%% file: tex/main_alg.tex
\algnewcommand{\LineComment}[1]{\State \(\triangleright\) #1}
\begin{algorithm}[t!]
\caption{Pseudocode for \algname. \algname proceeds in two stages. It first
estimates $p_k$ and $\sigma_k$. It then samples according to the estimated
optimal allocation, $\hat{T}_k = \sqrt{\hat{p}_k} \hat{\sigma}_k / \sum_{i=1}^K \sqrt{\hat{p}_i} \hat{\sigma}_i$.}
\label{alg:full-alg}
\begin{algorithmic}[1]
\Function{\textsc{ABaeInit}}{$\mathcal{D}$, $\mathcal{P}$, $K$}
    \State $\mathcal{D} \gets Sort(\mathcal{D}, \ key=lambda \ x : \mathcal{P}(x))$
    \State $\mathcal{S}_1, ..., \mathcal{S}_K \gets \text{StratifyByQuantile}(\mathcal{D}, K)$ 
    \State \Return $\mathcal{S}$
\EndFunction
\\
\Function{\textsc{ABaeSample}}{$\mathcal{S}$, $\mathcal{O}$, $K$, $N_1$, $N_2$, $\text{SampleFn}$}
    \For{each k in  [1, ..., K]} \Comment{Stage 1}
        \State $R_k^{(1)} \gets \text{SampleFn}(\mathcal{S}_k, N_1)$ \Comment{$R_k$ are sampled records}
        \State $X_k^{(1)} \gets \{f(x) \ | \ x \in R_k^{(1)}, \ \mathcal{O}(x) = 1\}$ 
        \State $\hat{\mu}_k$ $\gets$ $\sum_{i=1}^{ |X_k^{(1)}| } X_{k, i}^{(1)} / |X_k^{(1)}|$ if $|X_k^{(1)}| > 0$ else $0$
        \State $\hat{p}_k$ $\gets$ $|X_k^{(1)}| / |R_k^{(1)}|$
        \State $\hat{\sigma}^2_k$ $\gets$ $\sum_{i=1}^{ |X_k^{(1)}| } \frac{(X_{k, i}^{(1)} - \hat{\mu}_k)^2}{|X_k^{(1)}| - 1}$ if $|X_k^{(1)}| > 1$ else $0$
    \EndFor
    \For{each k in  [1, ..., K]} 
        \State $\hat{T}_k$ $\gets$ $\sqrt{\hat{p}_k} \hat{\sigma}_k / \sum_{i=1}^K \sqrt{\hat{p}_i} \hat{\sigma}_i$ 
    \EndFor
    \For{each k in  [1, ..., K]} \Comment{Stage 2}
        \State $R_k^{(2)} \gets R_k^{(1)} + \text{SampleFn}(\mathcal{S}_k, \floor{N_2 \hat{T}_k})$
        \State $X_k^{(2)}\gets X_k^{(1)} + \{f(x) \ | \ x \not \in R_k^{(1)}, \ x \in R_k^{(2)}, \ \mathcal{O}(x)=1\}$
        \State $\hat{p}_k$ $\gets$ $|X_k^{(2)}| / |R_k^{(2)}|$
        \State $\hat{\mu}_k$ $\gets$ $\sum_{i=1}^{|X_k^{(2)}|} X_{k, i}^{(2)} / |X_k^{(2)}| $ if $|X_k^{(2)}| > 0$ else $0$
    \EndFor
    \State \Return $\sum_{k=1}^K \hat{p}_k \hat{\mu}_k / \sum_{k=1}^K \hat{p}_k$, $R^{(2)}$
\EndFunction
\\
\Function{\algname}{$\mathcal{D}$, $\mathcal{O}$, $\mathcal{P}$, $K$, $N_1$, $N_2$}
    \State $\mathcal{S} \gets \textsc{ABaeInit}(\mathcal{D}, \mathcal{P}, K)$
    \State $\text{SampleFn} \gets \text{SampleWithoutReplacement}$
    \State $\hat{\mu}, \ R^{(2)} \gets \textsc{ABaeSample}(\mathcal{S}, \mathcal{O}, K, N_1, N_2, \text{SampleFn})$
    \State \Return $\hat{\mu}$
\EndFunction
\end{algorithmic}
\end{algorithm}

%% file: tex/bootstrap_alg.tex
\begin{algorithm}[t!]
    \caption{Bootstrap procedure for computing CIs. We resample existing samples over both stages of the algorithm.}
    \label{alg:bootstrap-alg}
    \begin{algorithmic}[1]
    \Function{Bootstrap}{$R^{(2)}$, $\mathcal{O}$, $K$, $N_1$, $N_2$, $\beta$, $\alpha$} 
        \For{each b in [1, ..., $\beta$]} \Comment{$\beta$ is \# of bootstrap trials}
            \For{each k in [1, ..., $K$]}
                \State $R_k^{*} \gets \text{SampleWithReplacement}(R_k^{(2)}, |R_k^{(2)}|)$
                \State $X_k^{*} \gets \{f(x) \ | \ x \in R_k^{*}, \ \mathcal{O}(x) = 1\}$
                \State $\hat{p}_k^* \gets |X_k^{*}| / |R_k^{*}|$
                \State $\hat{\mu}_k^* \gets \sum_{i=1}^{|X_k^*|} X_{k, i}^* / |X_k^*| $ if $|X_k^*| > 0$ else $0$
            \EndFor
            \State $\hat{\mu}_b \gets \sum_{k=1}^K \hat{p}_k^* \hat{\mu}_k^* / \sum_{k=1}^K \hat{p}_k^*$
        \EndFor
        \State \Return $\text{Percentile}(\alpha/2, \ \hat{\mu}), \text{Percentile}(1-\alpha/2, \ \hat{\mu})$
    \EndFunction
    \\
    \Function{ABaeWithCI}{$\mathcal{D}$, $\mathcal{O}$, $\mathcal{P}$, $K$, $N_1$, $N_2$, $\beta$, $\alpha$}
        \State $\mathcal{S} \gets \textsc{ABaeInit}(\mathcal{D}, \mathcal{P}, K)$
        \State $\text{SampleFn} \gets \text{SampleWithoutReplacement}$
        \State $\hat{\mu}, \ R^{(2)} \gets \textsc{ABaeSample}(\mathcal{S}, \mathcal{O}, K, N_1, N_2, \text{SampleFn})$
        \State \Return $\hat{\mu}, \ \textsc{Bootstrap}(R^{(2)}, \mathcal{O}, K, N_1, N_2, \beta, \alpha)$
    \EndFunction
    \end{algorithmic}
\end{algorithm}

%% file: tex/analysis.tex
\section{Theoretical Analysis}
\label{sec:analysis}

We present a statistical analysis of \algname and its extensions. We first show
that a related sampling procedure achieves rate $O\left( \frac{1}{N} \right)$
assuming perfect knowledge of $p_k$ and $\sigma_k$. We then show that our
sampling procedure matches the rate of the optimal strategy. Finally, we show
that our optimization procedure for allocation for group by keys is optimal for
the deterministic setting.

We provide the intuition and theorem statements in this manuscript. We defer the
full proofs to an extended technical report~\cite{kang2021proof}.

\subsection{Notation and Preliminaries}

\minihead{Notation}
Recall the notation in Table~\ref{table:notation}. We emphasize that
$X_{k, i}$ is the $i$th \emph{positive} sample from stratum $k$, i.e., the $i$th
sample that satisfies the predicate. Furthermore, recall that $\mu_k$ is the
per-stratum mean, $\pall = \sum p_k$ be the sum of the $p_k$, and $\muall$ be the
overall mean. Finally, recall that $w_k = p_k / \pall$, the normalized predicate
positive rate, which corresponds to the weighting of $\mu_k$ to $\muall$.

\minihead{Assumptions and properties}
We assume $X_{k, i}$ is sub-Gaussian with nonzero standard deviation,
a standard assumption for stratified sampling \cite{carpentier2015adaptive}. Sums of sub-Gaussian variables
converge with quantitative rates and this assumption widely holds in practice.
In particular, centered, bounded random variables are sub-Gaussian. The
sub-Gaussian assumption gives the existence of universal constants such that
$\Exp[|X_{k, i}|] \leq C^{(\mu)}$ and $Var[X_{i, k}] \leq C^{(\sigma^2)}$.

We further assume that $\pall \geq C^\pall > 0$, which enforces that at least
one stratum has non-vanishing $p_k$.

\subsection{Optimal Allocation with Deterministic Draws}
We first analyze the setting where we assume perfect knowledge of $p_k$ and
$\sigma_k$ and that we receive a deterministic, per-stratum number of draws
given a sampling budget. Specifically, given a budget of $T_k N$ per stratum, we
assume that we receive $p_k T_k N$ samples, rounded up. We prove the optimal
allocation under a continuous relaxation and the rate when using this optimal
allocation.

\begin{prop}
\label{prop:opt-alloc-deterministic}
Suppose $p_k$ and $\sigma_k$ are known and we receive $B_k = p_k T_k N$ samples
per stratum (up to rounding effects). Then, the choice $T_k = T_k^*$ that
minimizes the MSE for the unbiased estimator $\hmuall = \sum_k p_k \hat{\mu}_k / \sum_k p_k$
is
\begin{align}
  T_k^* = \frac{\sqrt{p_k}\sigma_k}{\sum_{i=1}^K \sqrt{p_i}\sigma_i}
\end{align}
\end{prop}

\begin{prop}
\label{prop:opt-mse-deterministic}
Suppose the conditions in Proposition~\ref{prop:opt-alloc-deterministic} hold.
Then, the squared error under the allocation $T_k^*$ is
\begin{align}
  \Exp[(\hmuall - \muall)^2 | B_k = p_k T_k^* N] &= \sum_{k=1}^K  \frac{w_k^2 \sigma_k^2}{p_kT_k^*N} \\
  &= \frac{1}{N \pall^2} \cdot \left(\sum_{k=1}^K \sqrt{p_k}\sigma_k\right)^2
\end{align}
\end{prop}

Intuitively, these propositions say for deterministic draws, the optimal
allocation downweights the standard importance sampling allocation by a factor
of $\sqrt{p_k}$. The resulting MSE decreases linearly with respect to the sample
budget and a scaling factor.

We note that uniform sampling with deterministic draws converges at rate
$\frac{\sigma^2}{N p_\textrm{avg}}$, where $p_\textrm{avg} = \sum p_k / K$. As a
result, stratified sampling offers room for improvement. For example, suppose
$p_1 = 1$, $p_k = 0$ for $k \neq 1$, and that $\sigma_k = 1$ for all $k$. This
corresponds to a perfect proxy and conditionally independent draws and
statistic. Then, uniform sampling converges at rate $\frac{K}{N}$, in contrast
to stratified sampling's rate of $\frac{1}{N}$. This corresponds to a $K$-fold
improvement in rate.

\subsection{\algname with a Single Predicate}
We analyze \algname's two stage sampling algorithm, in which we do not know
$p_k$ and $\sigma_k$. We provide the theorem statement, but defer the full proof
to an extended technical report~\cite{kang2021proof}. We assume that $N_2$ is suitably
large relative to $N_1$ for the remainder of the paper.

\begin{theorem}
\label{thm:abae-main}

With high probability over the draws made in Stage 1 and in expectation in Stage 2,
\begin{align}
  \Exp[(\hmuall - \muall)^2] \leq O\left(\frac{1}{N_1} + \frac{1}{N_2} + \frac{\sqrt{N_1}}{\sqrt{N_2}} \cdot \frac{1}{N_2}\right)
\end{align}
Furthermore, if $N_1 = N_2$
\begin{align}
  \Exp[(\hmuall - \muall)^2] \leq O\left(\frac{1}{N}\right)
\end{align}
\end{theorem}

\subsection{Understanding \algname}
We provide an overall proof sketch of the analysis of \algname and highlight
several aspects of the analysis of broader interest.

\subsubsection{Proof Sketch}
Our proof strategy proceeds as follows. We first show that our estimates
$\hat{p}_k$ and $\hat{\sigma}_k$ converge to $p_k$ and $\sigma_k$ in a
quantitative way (i.e., with a specific rate). As a result, our estimate for the
optimal allocation will also converge in a quantitative way.

Given the estimate for the optimal allocation, we show that the number of draws
in Stage 2 for all strata will approach the deterministic number of draws, for
$p_k$ large enough (larger than $\frac{1}{N_2}$). We then show that the error
converges appropriately for the strata with $p_k$ large enough and that the
error for the remaining strata becomes negligible. As a result, our final
estimate converges with rate $O(\frac{1}{N})$.

\subsubsection{Challenges}
We describe several challenges in the analysis of \algname. Prior work
has focused on known, deterministic per-strata costs and variances. In contrast,
our problem does not have a cost, but rather a stochastic probability of
receiving useful information. We study this stochastic draw case and prove that
using pilot sampling with plug-in estimates \cite{kish1965survey} is valid and near optimal.

\minihead{Unknown $p_k$ and $\sigma_k$}
Most work in stratified sampling assumes that features of the data distributions
within each stratum are known and constructs optimal allocations of samples
using this information. In our setting, these quantities must be estimated,
which may not be possible when $p_k$ is small. For example, if $p_k =
\frac{1}{N^2}$ for some stratum, then we may not draw even a single positive
record from that stratum, making $p_k$ and $\sigma_k$ impossible to estimate.

\minihead{Stochastic draws}
In contrast to standard stratified sampling, we may draw a record that does not
satisfy the predicate. As a result, for a fixed number of draws, the number of
records matching a predicate is stochastic. Most work on stratified sampling
assumes a deterministic allocation of samples to strata.

When the number of draws for some arbitrary $M$ from a stratum and the
probability $p_k$ of matching the predicate are both large, the
number of positive records concentrates around $p_k M$ and the resulting
estimator has similar properties to one with $p_k M$ deterministic draws.
However, if $p_k M$ is small, this analysis breaks down.

\minihead{Fractional allocations}
To show that \algname converges at the optimal rate, we compare to
the setting of deterministic draws and perfect information. Given perfect
information of $p_k$ and $\sigma_k$, the optimal allocation is given by
Proposition~\ref{prop:opt-alloc-deterministic} and its MSE is
given by Proposition~\ref{prop:opt-mse-deterministic}. However, this allocation
cannot be achieved in general, as it results in fractional sampling.
Nonetheless, we show that our sampling procedure, which rounds down the ideal
fractional allocations, achieves the same $O\left( \frac{1}{N} \right)$ rate.
Thus, rounding does not affect the convergence rate of our procedure.

\subsubsection{Statistical Intuition}
Our primary tool for dealing with unknown quantities and stochastic draws is
dividing the strata into groups: where $p_k$ is large and where $p_k$
is small. Since the number of positive draws is Binomial, we apply standard
convergence to the total number of positive draws when $p_k$ is large. For
stratum where $p_k$ is small, the contribution of that stratum to the total
error is at most $p_k C^{(\mu)}$, which does not increase the asymptotic error.
To illustrate our technique, consider the following proposition.

\begin{prop}
Recall that $N_1$ and $N_2$ are the number of samples in Stages 1 and 2
respectively. With high probability in Stage 1 and if $N_1$ is a constant
multiple of $N_2$ as $N$ grows, the MSE of the error in Stage 2 can be written
as
\begin{align}
\Exp&\left[ (\hmuall - \muall)^2 \right]
= \sum_{k=1}^K \hat{w}_k^2 \Var(\hat{\mu}_k) +
    O\left( \frac{1}{N_1} + \frac{1}{N_2} \right)
   \label{eq:mse-decomp}
\end{align}
where $\hat{w}_k = \hat{p}_k / \sum \hat{p}_k$.
\end{prop}

As shown in Eq.~\ref{eq:mse-decomp}, the overall MSE is bounded above by
the sum of $\hat{w}_k^2 \Var(\hat{\mu}_k)$, which are per-strata quantities. We
then bound these quantities for strata where
$p_k$ is (quantitatively) large or small. Specifically, define $p_* =
\frac{2 \ln(1 / \delta) + 2 \sqrt{\ln(1 / \delta)} + 2}{N_1}
= O\left( \frac{1}{N_1} \right)$
for failure probability $\delta$. Furthermore, we assume that $N_1$ is a
constant multiple of $N_2$ as $N$ grows. We divide the strata into cases based
on whether $p_k > p_*$ or $p_k \leq p_*$

Consider the case where $p_k > p_*$. By standard concentration arguments,
the number of positive samples in Stage 2 concentrates to its expectation,
which is large. Thus, $\hat{w}_k^2
\Var(\hat{\mu}_k)$ decays at rate (approximately) $O\left( \frac{1}{N_2}
\right)$ by standard concentration arguments. For $p_k \leq p_*$, we can
directly bound the contribution. To understand this, consider the following
proposition.

\begin{prop}
\begin{align}
\hat{w}_k^2 \Var[\hat{\mu}_k] &\leq \hat{w}_k^2 \left(
  \Exp \left[ \frac{\sigma_k^2}{B_k^{(2)}} | B_k^{(2)} > 0 \right] +
  P(B_k^{(2)} = 0)\mu_k^2
\right) \\
\label{eq:var-mu-k}
&\leq O \left( \frac{1}{N_1} + \frac{1}{N_2} + \frac{\sqrt{N_1}}{\sqrt{N_2}} \cdot \frac{1}{N_2} \right)
\end{align}
where $B_k^{(2)}$ is the number of positive draws in Stage 2.
\end{prop}

\begin{proofsketch}
The key challenge is bounding quantities involving $B_k^{(2)}$.
Suppose counterfactually that $B_k^{(2)}$ were deterministic: then the
expression would correspond to the standard variance of an i.i.d.~estimator.
Namely, the variance if an i.i.d.~estimator decays as $1/B_k^{(2)}$. However,
since we obtain a stochastic number of draws, we must condition on the event of
non-zero draws and take the expectation. Since the draws are binomial in
distribution, the leading order converges a.s.~to its mean value, which
would give the desired bound in this toy setting.

We now adapt this strategy to account for $B_k^{(2)}$ being stochastic in
\algname. For $p_k > p_*$, $B_k^{(2)}$ is approximately $p_k T_k^* N_2$ with
high probability. As a result, with high probability, each stratum had
sufficient samples to form estimates. We can complete the proof similarly to the
toy setting with deterministic $B_k^{(2)}$.

However, if $p_k < p_*$, we may not draw the requisite number of samples. For
example, if $p_k = \frac{1}{N^2}$, we would not obtain any samples on average.
Thus, our analysis must consider the case where $p_k < p_*$ separately. When
$p_k$ is small, we can directly bound the contribution of the sum.
Namely, $\hat{w}_k^2 = O(1/N^2)$ as $p_k \leq \frac{C_1}{N}$ and the remainder
of the quantities are bounded by a constant.


Thus, the overall bound follows from considering the strata where $p_k$ is small
and where $p_k$ is large.
\end{proofsketch}


\subsubsection{Overview of Techniques}
We briefly describe two techniques used to prove the necessary bounds.
First, we leverage exponential tail bounds on sums of Bernoulli random
variables (Lemma 1~\cite{chung2002connected}). Both the upper and lower tail
bounds are requires to show that $\sqrt{\hat{p}_k}$ converges to $\sqrt{p_k}$:
this requires stronger tail bounds than showing $\hat{p}_k$ converges to $p_k$.
Second, we use quantitative, exponential tail bounds on Binomial random
variables~\cite{tarjan2009chernoff} to bound the number of positive draws.

\subsection{Analyzing Group Bys}
\label{sec:group-by-analysis}
Recall that we aim to optimize the minimax error for queries with a group by
clause. Suppose there are $G$ groups and that we have a proxy per group. As in
the case with a single predicate, each proxy induces a stratification over the
dataset. Given these $G$ stratifications, \alggroup estimates the quantities
necessary for optimal allocation and executes Stage 2 of \algname on each
stratification appropriately. Thus, the key question is how to allocate samples
between the stratifications. We first demonstrate how to allocate samples in the
perfect information, deterministic setting and use our plug-in estimates for
allocation estimation.

For this section, we index the stratification by $l$, the group by $g$, and the
strata by $k$. Thus, $p_{l, g, k}$, $\sigma_{l, g, k}$, and $\mu_{l, g, k}$
denote the predicate positive rate, the standard deviation, and the stratum mean
in stratification $l$, group $g$, and strata $k$ respectively.

To accelerate group by queries, we rely on \algname as a subroutine. When we say
that we execute an instance of \algname with a stratification $l$, we mean that
we stratify the dataset using the proxy for group $l$ and that we allocate
samples across strata optimally for computing the statistic associated with that
group.

We now analyze the two cases described in Section~\ref{sec:groupby-desc}.

\minihead{Single Oracle}
In this scenario, recall that we can identify the group key with a single oracle
model. To accelerate this query, we will execute $G$ instances of \algname's
Stage 2 for each stratification $l$.

Since a single oracle model identifies the group key, applying \algname's
allocation for a given group gives us estimates for the other groups for
free. As a result, we can reuse these estimates across all groups to obtain
combined estimators. Thus, by uniformly sampling in Stage 1, \alggroup can
obtain estimates $\hat{p}_{l, g, k}$, $\hat{\sigma}_{l, g, k}$, and
$\hat{\mu}_{l, g, k}$. Given these quantities, we can estimate the error using
Proposition~\ref{prop:opt-mse-deterministic}. To account for the multiple
estimators across stratifications, we aggregate estimates via inverse-variance
weighting, which minimizes the variance~\cite{hartung2011statistical}.

Given the estimates from Stage 1, we estimate the optimal allocation across
stratifications with the following objective, where we allocate $\Lambda_l \cdot
N_2$ samples to stratification $l$:
%
\begin{align}
  \mathcal{L} = \min_{\Lambda \in [0, 1]^G, \ \sum_{l=1}^G \Lambda_l = 1} \left(
    \max_{g} \left(
      \sum_{l=1}^G \left( \frac{1}{\Lambda_l N_2}
        \sum_{k=1}^K \frac{\hat{w}_{l, g, k}^2 \hat{\sigma}_{l, g, k}^2} {\hat{p}_{l, g, k} \hat{T}_{l, k}}
        \right)^{-1}
    \right)^{-1}
  \right)
  \label{eq:group-multi}
\end{align}
The constraint that $\sum_{l=1}^G \Lambda_l = 1$ ensures at most $N_2$ samples
are used in Stage 2. The term in the inner sum is the estimated MSE using our
plug-in estimates.

This objective follows from the per stratification and per group error which can
be calculated using Proposition~\ref{prop:opt-mse-deterministic} and using that
inverse-variance weighting achieves the least error among all weighted averages
which can be calculated as $\left( \sum_i 1/\sigma_i^2 \right)^{-1}$.

Standard tools in convex optimization~\cite{boyd2004convex} show that the
objective and constraints are convex. Thus, the optimization problem has a
unique minimizer. We use the Nelder-Mead simplex algorithm
\cite{sorensen1982newton} to numerically compute the minimizer.

\minihead{Multiple Oracles}
In this scenario, determining which group, if any, a record belongs to requires
$G$ oracle models (one for each group), which we assume has similar costs. In
contrast to the setting with a single oracle model, we do not obtain estimates
for other groups when sampling a given group.

As we cannot obtain estimates for other groups in this setting, the oracle for
group $g$ is only applied to samples from stratification $g$. Namely, we only
consider elements where $g = l$. Specifically, we need only estimate
$\hat{p}_{g, g, k}$, $\hat{\sigma}_{g, g, k}$, and $\hat{\mu}_{g, g. k}$. Hence,
we will have a single final estimate $\hat{\mu}_{all, g, g}$ for each group $g$.
Using Proposition~\ref{prop:opt-mse-deterministic} we can estimate the error of
$\hat{\mu}_{all, g, g}$ as function of the number of samples we allocate towards
the instance of $\algname$ associated with stratification and group $g$. 

We allocate $N_1$ samples in Stage 1 for each group. To optimize for the
minimax error, we will Stage 2 for stratification $l$ with $\Lambda_l \cdot
N_2$ samples such that $\sum_{l=1}^G \Lambda_l = 1$. We will optimize for the
optimal values of $\Lambda$ with the following objective:
\begin{align}
  \mathcal{L} = \min_{\Lambda \in [0, 1]^G, \sum_{l=1}^G \Lambda_l = 1}
  \left(
    \max_{g} \frac{1}{\Lambda_l N_2} \sum_{k=1}^K \frac{\hat{w}_{g, g, k}^2 \hat{\sigma}_{g, g, k}^2}{\hat{p}_{g, g, k} \hat{T}_{g, k}}
  \right)
  \label{eq:group-single}
\end{align}
This objective follows from the formula for estimating error provided in
Proposition~\ref{prop:opt-mse-deterministic}. We further note that the objective
in Equation~\ref{eq:group-multi} reduces to the objective in
Equation~\ref{eq:group-single} when $p_{l, g, k} = 1, \sigma_{l, g, k} = \infty$
for $l \neq g$.

As before, we can use standard tools in convex optimization to show that the
objective and constraints are convex, so the optimization problem has a unique
minimizer. We also use the Nelder-Mead simplex algorithm to find the minimizer.

Asymptotic optimality of this objective follows from
Theorem~\ref{thm:abae-main}. As the per-group objectives are asymptotically
optimal, we can apply the union bound across the $G$ groups, which shows the
convergence of the overall objective.

\subsection{Discussion}
We have shown that \algname achieves the same asymptotic rate as the optimal
allocation strategy for deterministic draws. To our knowledge, the setting of
stochastic draws and our convergence proofs are novel. However, we defer
extensions to future work. For example, our analysis is asymptotic: finite
sample bounds with exact constants would compare against uniform sampling more
precisely. Additionally, a bandit algorithm that updates the estimates of $p_k$
and $\sigma_k$ per sample draw may provide non-asymptotic improvements.

\input{tex/eval-datasets.tex}

%% file: tex/eval-datasets.tex


\begin{table*}[t!]
\centering

\caption{Summary of datasets, predicates, target DNNs, and proxies.}
\label{table:datasets}

\small
\begin{tabular}{lllll}
  Dataset & Size & Predicate &  Target DNN & Proxy model \\ \hline
  \texttt{night-street} & 973,136   & At least one car & Mask R-CNN~\cite{he2017mask} & TASTI~\cite{kang2020task} \\
  \texttt{taipei}       & 1,187,850 & At least one car & Mask R-CNN~\cite{he2017mask} & TASTI~\cite{kang2020task} \\
  \texttt{celeba}~\cite{liu2015faceattributes}
      & 202,599 & Blonde hair & Human labels & MobileNetV2~\cite{sandler2018mobilenetv2} \\
  Amazon movie posters~\cite{ni2019justifying}
      & 35,815 & Contains woman & MT-CNN~\cite{zhang2016joint}, VGGFace~\cite{serengil2020lightface} & MobileNetV2~\cite{sandler2018mobilenetv2} \\
  \texttt{trec05p}~\cite{cormack2005trec}
      & 52,578 & Is spam & Human labels & Keyword-based \\
  Amazon office supplies~\cite{ni2019justifying} &
      800,144 & Strong positive sentiment & FlairNLP BERT sentiment~\cite{akbik2018coling} & NLTK sentiment~\cite{hutto2014vader}
\end{tabular}
\end{table*}

%% file: tex/eval.tex
\section{Evaluation}
\label{sec:eval}

We evaluate \algname and its extensions on six real world datasets and synthetic
datasets. We first describe the experimental setup and baselines. We then
demonstrate that \algname outperforms baselines in all settings we consider. We
also show that \algname's sample reuse is effective and \algname is not
sensitive to hyperparameters.

\subsection{Experimental Setup}
\minihead{Datasets, target DNNs, and proxies}
We consider six real world datasets, including text, still images, and videos
(Table~\ref{table:datasets}). We additionally consider synthetic datasets for
some settings.

We used the \texttt{night-street} (also known as \texttt{jackson}) and
\texttt{taipei} video datasets, which are commonly used for video analytics
evaluation~\cite{kang2017noscope, kang2019blazeit, canel2019scaling,
xu2019vstore, jiang2018chameleon}. We executed the following query:
\begin{lstlisting}
SELECT AVG(count_cars(frame)) FROM video
WHERE count_cars(frame) > 0
\end{lstlisting}
which computes the average number of cars in the video, conditioning on cars
present. We use Mask R-CNN to compute the oracle filter \cite{he2017mask}.
We use an efficient index for the proxy scores \cite{kang2020task}.

We used the \texttt{celeba} dataset~\cite{liu2015faceattributes}, an image
dataset of celebrity faces that contains annotations of celebrity names and
other attributes, such as hair color. We executed the following query:
\begin{lstlisting}
SELECT PERCENTAGE(is_smiling(img)) FROM images
WHERE hair_color(img) = 'blonde'
\end{lstlisting}
which computes the fraction of images where the celebrity is smiling conditioned
on the celebrity having blonde hair. We used the human labels in the
\texttt{celeba} dataset as the ground truth. We used a specialized
MobileNetV2~\cite{sandler2018mobilenetv2} as the proxy.

We used the TREC public spam corpora from 2005
(\texttt{trec05p})~\cite{cormack2005trec}. We used the SPAM25 subset. We
executed the following query:
\begin{lstlisting}
SELECT AVG(NB_LINKS(text)) FROM emails
WHERE is_spam(text)
\end{lstlisting}
which computes the average number of links for spam emails. We used human labels
as ground truth. We used a manual, keyword-based proxy based on the presence of
words (e.g., ``money,'' ``please'').

We used Amazon movie reviews and posters, which was generated from the Amazon
reviews dataset~\cite{ni2019justifying}. We scraped the movie posters from the
metadata and excluded reviews that did not have posters. We executed the
following query:
\begin{lstlisting}
SELECT AVG(rating) FROM movies
WHERE face_exists(poster) AND gender(poster) = 'female'
\end{lstlisting}
which computes the average rating of posters with a female actress. We use
MT-CNN to extract faces~\cite{zhang2016joint} and VGGFace pretrained from
deepface~\cite{serengil2020lightface} to classifier gender as the ground truth.
We use a specialized MobileNetV2 as a proxy~\cite{sandler2018mobilenetv2}.

We used the Amazon reviews dataset~\cite{ni2019justifying} which is a dataset of
textual reviews from Amazon. We subset to the office supplies reviews. We
executed the following query:
\begin{lstlisting}
SELECT AVG(rating) FROM data
WHERE sentiment(review) = 'strongly positive'
\end{lstlisting}
which computes the average rating of reviews with strongly positive sentiment.
We use a BERT-based sentiment classifier provided by FlairNLP to compute the
oracle filter~\cite{akbik2018coling} and the NLTK sentiment predictor, a simple
rule-based classifier, for the proxy~\cite{hutto2014vader}.

\minihead{Metrics}
Our primary metric is the \emph{RMSE} of the true and estimated
values: we use the RMSE so that the units are on the same scale as the original
value. We additionally compare the number of samples required to achieve a
particular error target in some experiments.
We measure the cost in terms of oracle predicate invocations as it is the
dominant cost of query execution by orders of magnitude.

\minihead{Methods evaluated}
We compare \algname to uniform sampling as it is applicable without precomputing predicate results. A range of standard AQP
techniques are not applicable to our setting, since the results of the predicate
are not available at ingest time. For example, techniques that create
histograms~\cite{piatetsky1984accurate, poosala1996improved,
cormode2009probabilistic} or sketches~\cite{garofalakis2002querying,
gan2020coopstore} as ingest time are not applicable.

\minihead{Implementation}
We implement \algname's sampling procedure in Python for ease of integration
with deep learning frameworks.
Our
open-sourced code is available at
\url{https://github.com/stanford-futuredata/abae}.

\input{tex/eval-fu.tex}
\input{tex/eval-sensitivity.tex}

%% file: tex/eval-fu.tex
\subsection{End-to-end Performance}
\label{sec:eval-fu}

\begin{figure}[t!]
  \includegraphics[width=\columnwidth]{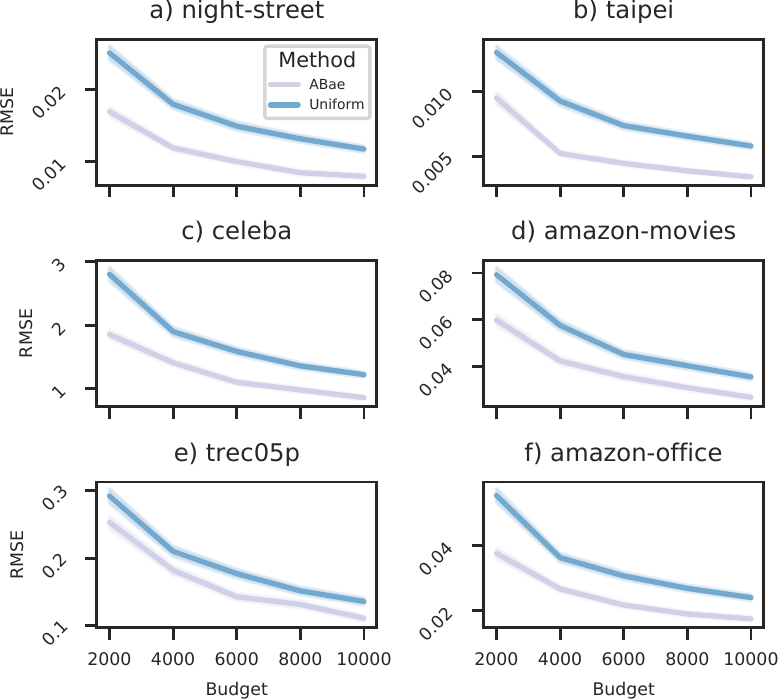}
  \caption{
  Sampling budget vs RMSE for uniform sampling and \algname, with
  the standard deviation shaded. \algname outperforms on all budgets
  and datasets we evaluated on. \algname can outperform by up to 1.5$\times$
  on RMSE at a fixed budget and achieve the same error with up to 2$\times$
  fewer samples.
  }
  \label{fig:fu-mse}
\end{figure}

\begin{figure}[t!]
  \includegraphics[width=\columnwidth]{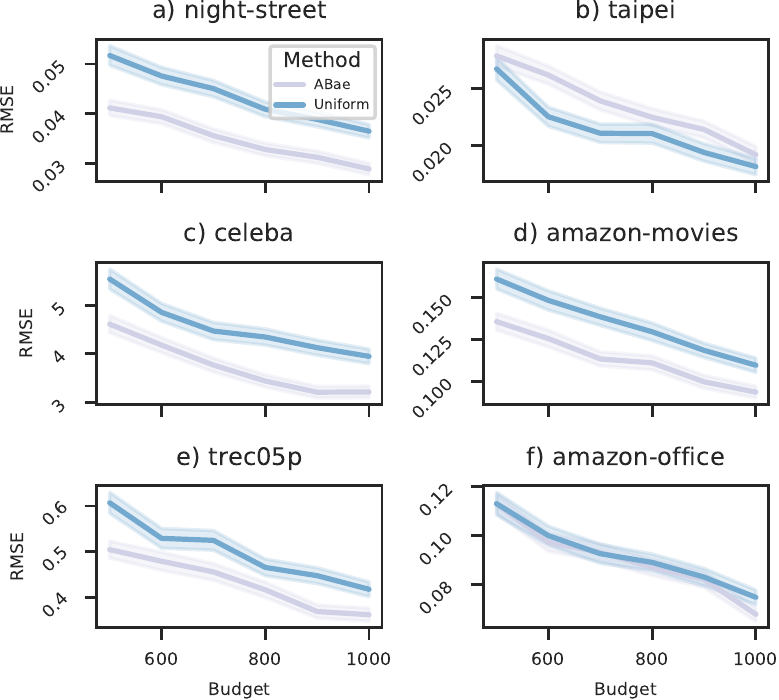}
  \caption{
    Low sampling budgets vs RMSE for uniform sampling and
    \algname, with the standard deviation shaded. We see that even at small
    sample sizes, \algname outperforms or matches uniform sampling in all
    cases.
    \label{fig:fu-mse-small}
  }
\end{figure}

\begin{figure}[t!]
  \includegraphics[width=\columnwidth]{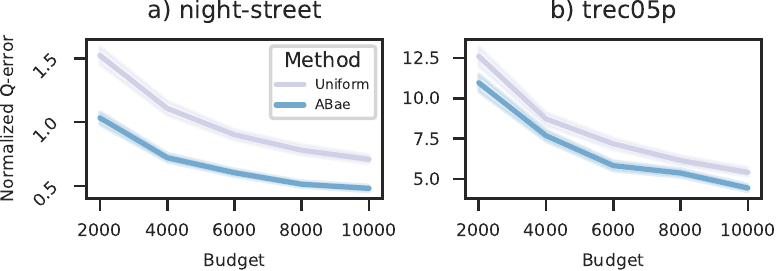}
  \caption{
    Sampling budget vs normalized Q-error for uniform sampling and \algname,
    with the standard deviation shaded. We see that \algname outperforms on
    Q-error. The same trends hold for all other datasets.
    \label{fig:q-error}
  }
\end{figure}

\begin{figure}[t!]
  \includegraphics[width=\columnwidth]{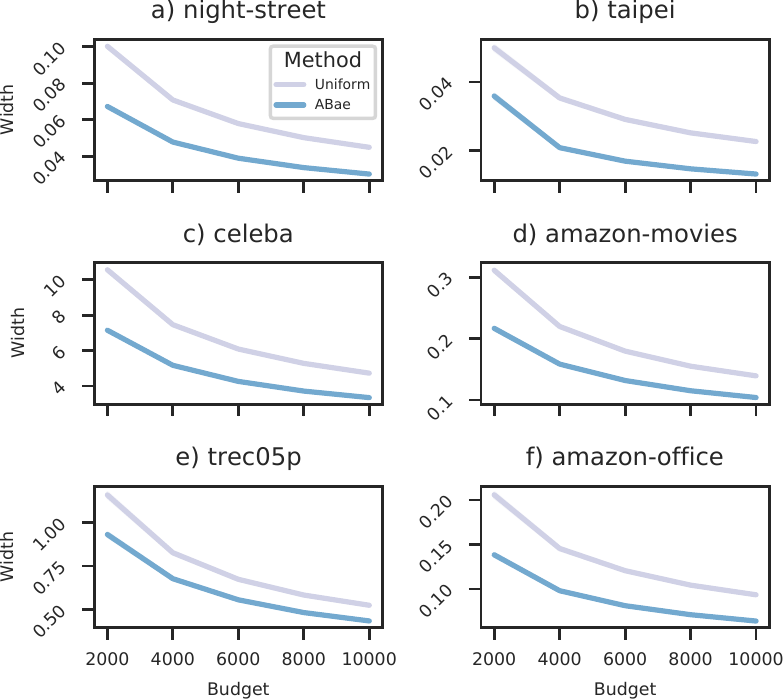}
  \caption{
  Sampling budget vs CI width for uniform sampling and \algname
  with the standard deviation of the width shaded (not visible in many plots).
  \algname can outperform by up to 1.5$\times$ on CI width at a
  fixed budget and achieve the same width with up to 2$\times$ fewer samples.
  }
  \label{fig:fu-ci}
\end{figure}

\minihead{Single predicate}
We show that \algname outperforms uniform sampling on the metric of RMSE. For
each dataset and query, we executed \algname and random sampling for sampling
budgets of 2,000 to 10,000 in increments of 2,000. We used five strata and
allocated half budget to each stage. We used a failure probability of 5\% for
every condition. We ran every condition 1,000 times.

As shown in Figure~\ref{fig:fu-mse}, \algname outperforms for every dataset,
query, and budget setting we consider. \algname can achieve up to 2.3$\times$
improvements in RMSE at a fixed budget or up to 2$\times$ fewer samples at a
fixed error rate. We additionally show that \algname outperforms uniform
sampling at low sampling budgets (500-1,000) in Figure~\ref{fig:fu-mse-small}.

We further show that \algname outperforms on
Q-error~\cite{moerkotte2009preventing}, which is a relative error metric that
penalizes under- and over-estimation symmetrically. We show the normalized
Q-error (i.e., $100\times (q - 1)$), which roughly indicates percent error in
Figure~\ref{fig:q-error}. As shown, \algname outperforms on the two datasets we
show--\algname also outperforms on all the other datasets by 14-70\%, which we
omit for brevity. \algname similarly outperform on relative error by 13-76\%.

We further show that \algname outperforms on the metric of confidence interval
(CI) width. For each dataset and query, we executed \algname and random sampling
with the parameters above. We ran every condition 1,000 times.

\algname outperforms for every dataset,
query, and budget setting we consider (Figure~\ref{fig:fu-ci}). \algname can outperform by up to
1.5$\times$ on CI width at a fixed budget. Furthermore, to
achieve the same confidence interval width, \algname can use up to 2$\times$
fewer samples.
Finally, \algname satisfies the nominal coverage across all
datasets and settings.

\begin{figure}[t!]
  \includegraphics[width=\columnwidth]{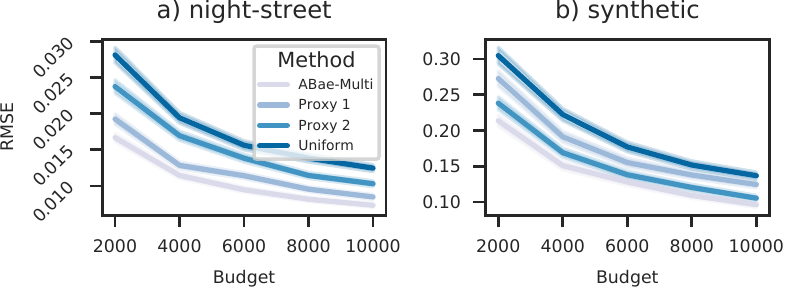}
  \caption{
  Sampling budget vs RMSE for uniform sampling and \algmulti, with
  the standard deviation shaded. As shown, \algmulti outperforms on both queries
  and all budgets we evaluated on.
  }
  \label{fig:complex}
\end{figure}

\minihead{Multiple predicates}
We show that \algmulti outperforms random sampling and using a proxy for a
single predicate. For the \texttt{night-street} dataset, we executed the
following query:
\begin{lstlisting}
SELECT AVG(count_cars(frame)) FROM video
WHERE count_cars(frame) > 0
  AND red_light(frame)
\end{lstlisting}
The positive rate is $0.17$. We additionally executed a query on a synthetic
dataset with five strata and two predicates. For each proxy, we draw the $p_k$
values from a Beta distribution. As shown in Figure~\ref{fig:complex}, \algmulti
outperforms on both queries and every budget setting we consider.


\begin{figure}[t!]
  \includegraphics[width=\columnwidth]{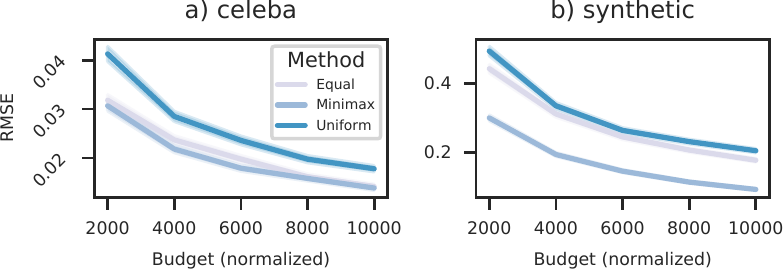}
  \caption{Normalized sampling budget vs max RMSE for uniform
  sampling and \alggroup with a single oracle (standard deviation shaded).
  \alggroup outperforms on both queries and on all budgets we evaluated on.}
  \label{fig:groupby-single}
\end{figure}

\begin{figure}[t!]
  \includegraphics[width=\columnwidth]{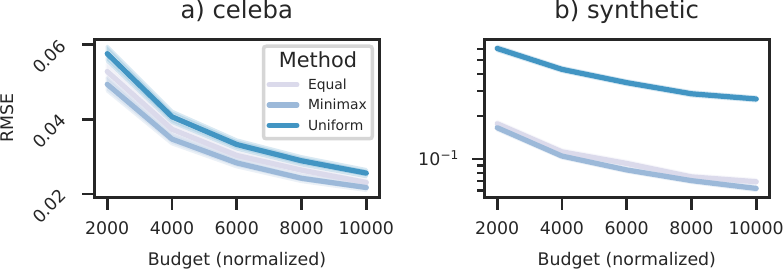}
  \caption{Normalized sampling budget vs max RMSE
  for uniform sampling and \alggroup with multiple, per-group oracle, with the
  standard deviation shaded. The y-axis is on a log-scale. The
  budget is normalized by the total number of groups. \alggroup
  outperforms on both queries and on all budgets we evaluated on.}
  \label{fig:groupby-multiple}
\end{figure}

\minihead{Group bys (single oracle)}
We show that \alggroup outperforms random sampling in the single oracle setting.
For the \texttt{celeba} dataset, we executed the following query:
\begin{lstlisting}
SELECT PERCENTAGE(is_smiling(image)) FROM images
WHERE HAIR_COLOR(image) = "gray"
   OR HAIR_COLOR(image) = "blond"
GROUP BY HAIR_COLOR(image)
\end{lstlisting}
We additionally executed a synthetic query where the statistic was
distributed normally and the predicate was generated as a Bernoulli with the
proxy probability. The synthetic dataset had four groups with a positive rate of
3.3\%, 3.3\%, 3.4\%, and 3.5\% respectively.

For these queries, we normalized the budget by the total number of groups. We
measured the maximum RMSE over all groups. As shown in
Figure~\ref{fig:groupby-single}, \alggroup outperforms on both queries and all
budget settings we consider.

\minihead{Group bys (multiple oracles)}
We show that \alggroup outperforms random sampling when a separate oracle is
requires for each group by key. For the \texttt{celeba} dataset, we executed the
same query as for group bys witn a single oracle. We additionally executed a
synthetic query where the statistic was distributed normally and the predicate
was generated as a Bernoulli with the proxy probability. The synthetic dataset
ahd four groups with positive rates of 16\%, 12\%, 9\%, and 5\% respectively.

For these queries, we normalized the budget by the total number of groups as
extracting the group key requires executing multiple models. We measured the
maximum RMSE over all groups. As shown in Figure~\ref{fig:groupby-multiple},
\alggroup outperforms on both queries and all budget settings we consider.

\minihead{Discussion of results}
To contextualize our results, we first note that relative errors for some of the datasets are as
high as 12\% (Figure~\ref{fig:q-error}b).
As a result, a 2$\times$ decrease in error (or number of samples at a fixed
error) represents a substantial improvement. Several of our ongoing
collaborations at Stanford University and elsewhere require expert human
labeling as part of scientific studies. Requiring 2$\times$ fewer human labels
is a substantial decrease in expert time.

%% file: tex/eval-sensitivity.tex
\subsection{Lesion and Sensitivity Analysis}
\label{sec:eval-sensitivity}

\begin{figure}[t!]
  \includegraphics[width=\columnwidth]{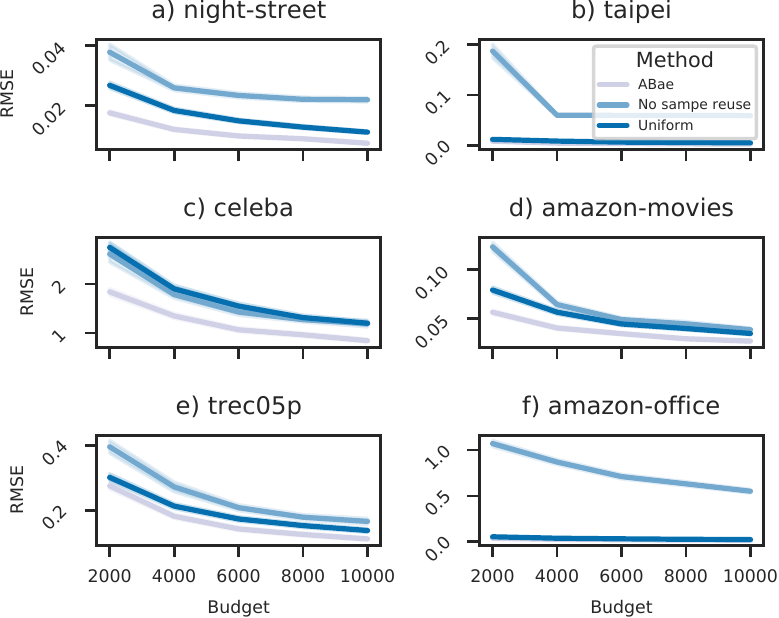}
  \caption{
  Lesion study when removing sample reuse and all components (i.e., uniform
  sampling) of \algname. As shown, both sample reuse and allocation between
  strata are critical for performance on all datasets.
  }
  \label{fig:lesion}
\end{figure}


\minihead{Lesion study}
We performed a lesion study in which we removed sample reuse and our two stage
procedure (i.e., uniform sampling). Specifically, we executed \algname, \algname
without sample reuse, and random sampling on all datasets. We used 10,000
samples for all conditions and ran 1,000 trials. For \algname, we used five
strata and allocated half of the samples in each stage.

As shown in Figure~\ref{fig:lesion},
both parts of \algname are necessary for high performance. In particular,
removing sample reuse substantially harms query performance. Sample reuse
contributes to more accurate estimates of $p_k$, which is critical for low
error.

\begin{figure}[t!]
  \includegraphics[width=\columnwidth]{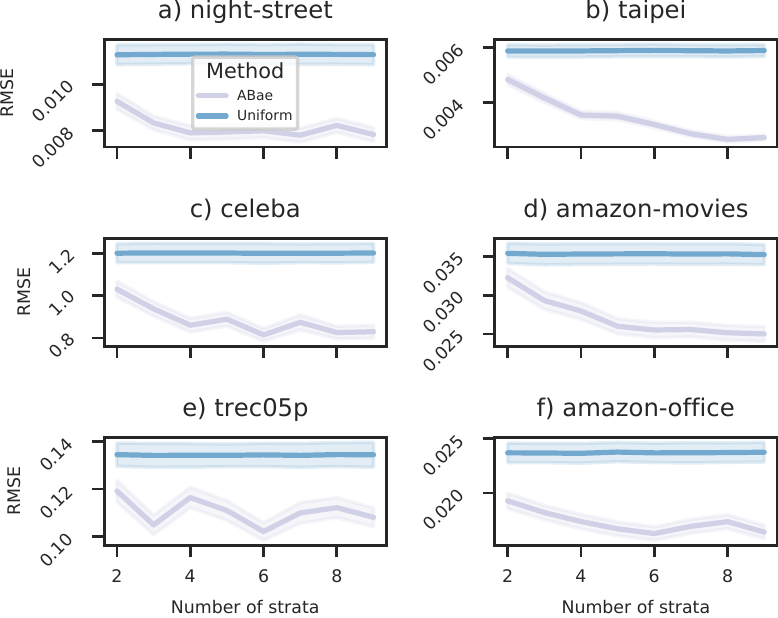}
  \caption{
    Sensitivity analysis of \algname to the number of strata. As shown, \algname
    outperforms on all values of $K$ we evaluated on compared to uniform sampling.
  }
  \label{fig:sensitivity-k}
\end{figure}

\begin{figure}[t!]
  \includegraphics[width=\columnwidth]{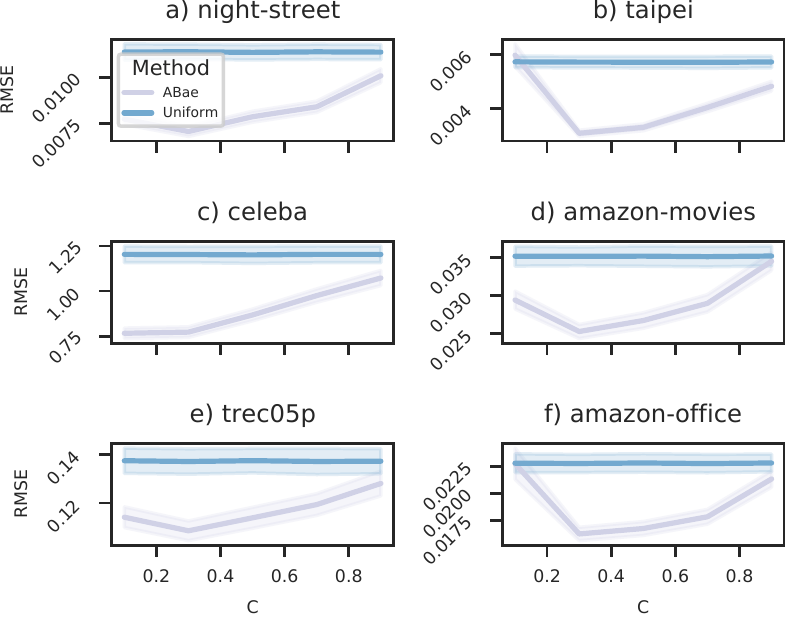}
  \caption{
    Sensitivity analysis of \algname to the fraction of samples between Stage
    1 and Stage 2. As shown, \algname generally outperforms, except for
    extreme values of $C$.
  }
  \label{fig:sensitivity-c}
\end{figure}

\minihead{Sensitivity analysis}
We analyzed the sensitivity of \algname to the number of strata ($K$) and the
fraction of samples between Stage 1 and Stage 2 ($C$). We executed \algname when
varying $K$ and $C$ with a budget of 10,000 and ran 1,000 trials for each
condition and dataset.

We varied $K$ from two to 10 and compared to uniform sampling. As shown in
Figure~\ref{fig:sensitivity-k}, \algname outperforms on all choices of $K$ for
all datasets. We find that, perhaps surprisingly, the number of strata does does
not affect strongly performance relative to the performance of uniform sampling.
However, in our datasets, more strata tends to perform better.

We varied $C$ from 0.1 to 0.9 in increments of 0.2. We compared to uniform
sampling. As shown in Figure~\ref{fig:sensitivity-c}, \algname outperforms for
$C$ between 0.3 and 0.7, a wide range of values.  Several datasets
underperform on extreme values of $C$ (0.1 and 0.9), but these are outside of
our recommended values of $C$.

\begin{figure}[t!]
  \includegraphics[width=\columnwidth]{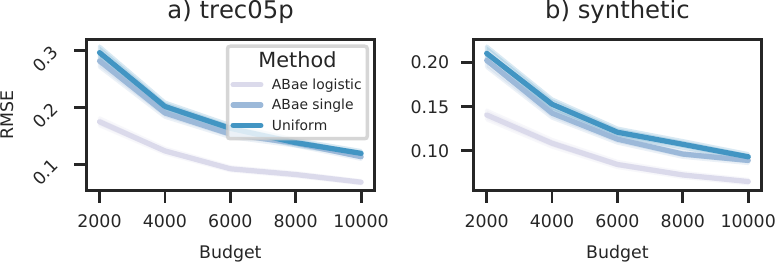}
  \caption{
  Uniform sampling, \algname with a single proxy, and \algname combining
  proxies. As shown, \algname's method of combining proxies outperforms
  baselines.
  }
  \label{fig:proxy-regression}
\end{figure}

\minihead{Combining proxies}
We analyzed if \algname's method of combining proxies via logistic regression can
improve performance. We used keyword-based proxies for the \texttt{trec05p}
dataset and a synthetic dataset. For the synthetic dataset, we generated
Bernoulli random variables and the proxies were the Bernoulli parameters with
noise.

We show results for uniform sampling, single proxy \algname , and \algname
with the combined proxy in Figure~\ref{fig:proxy-regression}. As shown, \algname
can combine proxies, effectively ``ignoring'' low quality proxies.

%% file: tex/rel_work.tex
\section{Related Work}
\label{sec:rel-work}

\minihead{AQP}
A recent survey on AQP techniques categorizes AQP into offline and online
methods~\cite{li2018approximate}. The closest work in AQP are offline methods,
which use pre-computed samples~\cite{agarwal2013blinkdb, acharya1999aqua},
histograms~\cite{piatetsky1984accurate, poosala1996improved,
cormode2009probabilistic}, wavelets~\cite{guha2005wavelet}, and
sketches~\cite{garofalakis2002querying, gan2020coopstore} to accelerate
approximate queries. These techniques involve computation at ingest time to
generate a synopses based on expected query workload~\cite{li2018approximate},
but assume the records are already present as structured data. Since we assume
the predicates are expensive to compute, we cannot compute these synopses at
ingest time.

\algname is also related to online methods, where the statistic is computed
on the fly. For example, online aggregation provides shrinking confidence
intervals as the query is executed~\cite{hellerstein1997online}. These
techniques also largely rely on precomputed information, e.g., indexes.

\minihead{Surveying and optimal allocation}
Work in the surveying and classical sampling literature have long studied
stratified sampling \cite{scheaffer2011elementary, nassiuma2001survey}.
If the
strata variances and costs are known, the optimal allocation is proportional to
$\frac{\sigma_k}{\sqrt{c_k}}$, for costs $c_k$ \cite{kish1965survey}.
The closest work we are aware of are algorithms for stratified sampling where
the strata variances are not known~\cite{carpentier2015adaptive,
arouna2004adaptative}. This work does not consider the case of stochastic draws,
as we do.

Other work in surveying considers stratified sampling with non-responses, such
as non-responses to mail surveys~\cite{hansen1946problem}. This literature is
largely concerned with estimating the bias from non-responses or focusing on
which subpopulations to follow up with, e.g., with phone
calls~\cite{khan2008optimum, filion1975estimating, varshney2012optimum}. In this
work, not all of the population satisfies the predicate, so the non-response
model is different.

A common technique in scientific studies and surveying is pilot
sampling, in which a small sample is used for preliminary analysis. Pilot
sampling is commonly used in randomized trials \cite{thabane2010tutorial} and to
estimate various quantities for downstream sampling (strata variances, sampling
costs, feasibility, etc.) \cite{kish1965survey,
denne1999estimating, eldridge2016defining, cochran2007sampling}. In our setting,
each sample has a stochastic probability of giving useful information instead of
a cost, which is not covered by prior work. However, the allocations in \algname
coincide with those made by defining hypothetical costs proportional to
$\frac{1}{p_k}$. To show that this allocation remains valid in our setting, we
prove that Stage 1 of \algname is optimal with high probability, something which
is not handled by standard survey sampling theory.



\minihead{Stratified sampling in data management systems}
Several traditional systems and algorithms use stratified sampling to accelerate
sampling~\cite{agarwal2013blinkdb, acharya1999aqua, chaudhuri2007optimized}. In
the data management literature, using stratified sampling typically requires
pre-computation, which is not applicable for the same reasons as described
above.

Recent work learns machine learning models, such as classification DNNs or
generative adversarial networks (GANs), over relational data to improve
AQP~\cite{li2020supporting, thirumuruganathan2020approximate,
walenz2019learning}. These approaches approximate the data (e.g., the result of
a predicate or overall data statistics), which can be faster than directly
accessing the data. However, they do not apply to complex unstructured data
sources that we consider in this work.

\minihead{Proxies}
Recent work has focused on accelerated DNN-based queries by using proxies. Many
systems have been developed to accelerate certain classes of queries using
proxies, including selection without statistical
guarantees~\cite{kang2017noscope, lu2018accelerating, anderson2018predicate,
hsieh2018focus}, selection with statistical
guarantees~\cite{kang2020approximate}, aggregation queries without
predicates~\cite{kang2019blazeit}, and limit queries~\cite{kang2019blazeit}.
The work on selection does not directly apply to our setting, since these
systems do not guarantee that the records selected are independent of the
statistic we wish to compute. The \blazeit system accelerates aggregation
queries over the entire dataset by using proxies as a control variate, but does
not consider aggregation queries with predicates~\cite{kang2019blazeit}. We add
to this body of literature by designing an algorithm for stochastic draws
to address that many aggregation queries contain predicates.

\minihead{DNN-based queries}
Other work aims to accelerate other DNN-based queries. Several systems aim to
improve execution speeds or latency of DNNs on accelerators~\cite{zhang2017live,
poms2018scanner, kang2021jointly}, which are complementary to our work. 
Other work assumes that the target DNN is not expensive to execute or that
extracting bounding boxes is not expensive~\cite{zhang2019panorama,
fu2019rekall}. We have found that many applications require accurate and
expensive target DNNs, so we focus on reducing executing the target DNN via
sampling.

%% file: tex/conclusion.tex
\section{Conclusion}
\label{sec:conclusion}

To reduce the cost of approximate aggregation queries with expensive predicates,
we introduce stratified sampling algorithms leveraging proxies. We provide
proofs of convergence for stratified sampling with stochastic draws, which
corresponds to our setting. We show that \algname achieves optimal rates. We
further extend \algname to answer queries with multiple predicates and group by
keys. We show that our algorithms outperform baselines by up to 2.3$\times$ on a
wide range of domains and predicates.